\documentclass[12pt]{article}
\usepackage{epsf}
\usepackage{latexsym, amssymb}

\begin{document}

\title{
Material failure time and the fiber bundle model with thermal
noise
\author{A. Guarino, R. Scorretti $^\ast$ and S. Ciliberto \\
\\ Laboratoire de Physique, C.N.R.S., UMR 5672  \\
  Ecole Normale Superieure de Lyon, 46 all\'{e}e d'Italie,
  \\  69364 Lyon,
  France}}
\date{}
\maketitle

\begin{abstract}

The statistical properties of failure are studied in a fiber
bundle model with thermal noise. We find that in agreement with
recent experiments the macroscopic failure is produced by a
thermal activation of microcracks. Most importantly the effective
temperature of the system is amplified by the spatial disorder
(heterogeneity) of the fiber bundle.

\end{abstract}

\bigskip


\newpage

Material failure is a widely studied phenomenon  not only
 for its very  important  technological
applications but also for its fundamental statistical aspects,
which are not yet very well understood. Many models  have been
proposed to give more insight into the statistical analysis of
failure. Among the most studied ones we can mention the fuse and
the non-linear spring networks
\cite{hermann,hemmer,andersen,vanneste,zapperi} which can
reproduce several features of crack precursors, experimentally
observed in heterogeneous materials
 subjected to a quasi-statically increasing  stress
 \cite{lockner,anifrani,noiprl,petri,maes}. Specifically the power law behaviour  of the
acoustic emission observed in several experiments close to the
failure point.

  However  these networks and  the other related    models, in their
  standard formulation,
   are unable to describe the behaviour of a material
subjected to a creep-test, which consists in keeping a sample at a
constant stress till  it  fails. Creep-tests are widely used by
engineers  in order to  estimate the sample life time as a
function of the applied stress.  A modified fuse network, which
takes into account the Joule effect in the fuses,  has been
 proposed to explain the finite life time of a sample
subjected to a constant stress\cite{vanneste}.
 However also this
model does not explain the recent experimental results on micro
crystals\cite{pauchard}, gels\cite{bonn} and heterogeneous
materials\cite{noi}. These experiments show that the life time
$\tau$ of a sample, subjected to an imposed stress $P$ is well
predicted by the equation

\begin{equation}
\tau =\tau _o \exp{\left( \alpha \frac{\Gamma
^dY^{(d-1)}}{KT_{eff} \ P^{(2d-2)}}\right) }  \label{const}
\end{equation}

where $\tau _o$ is a constant, $\Gamma $ the surface energy, $Y$
the Young modulus, $k$ the Boltzmann constant, $\alpha $ a
constant which depends on the geometry, $T_{eff}$ is an effective
temperature and $d$ the dimensionality of the system.  The main
physical hypothesis behind eq.\ref{const} is that the macroscopic
failure of a material is produced by a thermal activation of micro
cracks\cite{pomeau,golubovic}. In the original Pomeau's theory
\cite{pomeau} $T_{eff}$ of eq.\ref{const}  coincides with the
thermodynamic temperature $T$ while
 experimentally $T_{eff}>> T$\cite{pauchard,noi}.  To explain this results it has
 been supposed that  the disorder of the material "amplifies" the
 thermal noise\cite{noi}.

The purpose of this letter is to show that adding  to the spring
network  a noise, which plays the role of a  temperature,  it is
possible to reproduce the behaviour of a material subjected to a
creep-test.
 Specifically  the functional dependence on the applied
stress  of the network life time is consistent with that observed
in recent experiments, that is with  eq.\ref{const}. Furthermore
we can prove that, for such a process,   the noise is "amplified"
by the network disorder.

In order  to study the influence of noise on macroscopic failure
we have chosen  the simplest spring network, that is
 the so called
democratic fiber bundle model, proposed  a long time ago   by
Pierce \cite{pierce} to study cable failure. This model, widely
studied in the quasi static regime
\cite{hermann,hemmer,andersen,vanneste},  is equivalent to $N$
springs in parallel subjected to a total traction  force $F$.
Specifically we have numerically studied  the model by using the
following rules:

\begin{itemize}
\item[A1]  The external applied force $F$ produces a {\it local stress} $f_i$
on each fiber. $F$ is democratically  and completely distributed
in the net: $F=\sum_{i=1}^Nf_i$

\item[A2]  The local stress $f_i$ on the i-th fiber produces a {\it local
deformation} $e_i$. Being the media elastic, stress and
deformation are linked by Hook's law:
\begin{equation}
f_i=Y\cdot e_i
\end{equation}
where $Y$ is the Young modulus, which is assumed to be the same
for all the fibers : $Y_i=Y$.

\item[A3]  The strength of each fiber is characterized by a {\it critical
stress} $f_i^{(c)}$: if at time $t$ on the i-th fiber local stress
$f_i(t)$ is greater than critical stress $f_i^{(c)}$ the fiber
cracks, and his local stress falls to zero at time $t+1$. Further,
we assume that in this process some energy $\epsilon _i$ is
released proportionally to the square of local stress $f_i$; for
sake of simplicity we will assume $\epsilon _i=f_i^2$. Critical
stress $f_i^{(c)}$ is a realization of a random variable that
follows a normal distribution of mean $f^{(c)}$ and variance
$KT_d$:
\begin{equation}
f_i^{(c)}= f^{(c)}+ N_d(KT_d)
\end{equation}
We call $N_d$  {\it disorder noise}.

\item[A4]  Each fiber is subjected to an addictive time dependent
 random stress $\Delta
f_i(t)$ which follows a zero  mean normal distribution of variance
$KT$:
\begin{equation}
\Delta f_i(t)= N_T(t,KT)
\end{equation}
being $t$ the time. We  call $N_T$ {\it thermal noise}. We assume that $%
\Delta f_i(t)$ is a white random process,which is  independent in
each fiber, i.e.  the correlation function $E\left[ \Delta
f_i(t_1)\cdot \Delta f_j(t_2)\right] =0$ if $t_1\neq t_2$ or
$i\neq j$.

\end{itemize}

The first three items are those used in the standard formulation
of fiber bundle model. A  new feature,  which is similar to a
thermal activation process, is introduced in [A4] to explain the
dependence of failure time on a constant applied stress.

The model has the following properties.
  We see from [A1] that there exists a long-range
interaction between the fibers: indeed if a number $n(t)$ of
fibers are broken at time $t$, the local stress on each of the
remaining fibers will be:
\begin{equation}
f_i(t)=\frac F{N-n(t)}+\Delta f_i(t)
\end{equation}
that is, under the specified boundary conditions, the breaking of
some fibers produces an increase of local stress on the other
ones. Item [A3] models the heterogeneity of media. The assumption
that all the disorder in the model appears in the strength
distribution rather than in the elastic constants may be argued by
noticing that the effective elastic constant of a single fiber is
essentially the average of the local elastic constant along the
fiber, while the strength is determined by its weakest point
\cite{hemmer}. If $KT=0$ the model is reduced to the standard one:
in this case the applied force is  increased linearly  $F=A_p t$
from zero to the critical value $F_c$ needed to break the whole
network. The sample life time $\tau$ is  equal to $F_c/A_p$ for
any value of the slope $A_p$.

 It is quite
clear that when $KT=0$ and  a constant force $F$ is applied the
system  breaks in a single avalanche only if $F$ is large enough
otherwise it will never break. For example
 if $KT_d=0$ all the fibers are strictly
equal and if there is no thermal activation ( i.e. $KT=0$) the
system breaks in a single step  when  $F=N f^{(c)}$.

If $KT \ne 0$ then the system can break also at constant imposed
stress. We have numerically studied the behaviour of the model as
a function of $F$, $KT$ and $KT_d$. In the following we assume
$N=1000$ and $f^{(c)}=1$. For each set of parameter values  we
have repeated the numerical simulation at least ten times to
estimate the  scattering of the results for different realization
of the noise. This corresponds to the scattering of symbols in the
figures. We call {\it event} the simultaneous  breaking of several
fibers and {\it
 event size } the number $s(t)$ of fibers which crack. The energy
$\epsilon $ associated to an event is the sum of the energies
released by the fibers which crack, that is :
\begin{equation}
\epsilon =s(t)\cdot \left( \frac F{N-n(t)}\right) ^2.
\end{equation}
 The cumulated energy $E(t)$ is the sum of $\epsilon$ from 0 to
t.
 When $KT\ne0$ and $F$ is
constant the event  statistics  is quite similar to that observed
at  $KT=0$ for $F$ increasing linearly in time
\cite{hermann,hemmer,zapperi}.
 As an example of the system response  to a constant
force $F=540$ with $KT_d=0.005$ and $KT=0.007$, we plot in fig.1a
the distribution $N(\epsilon)$ of $\epsilon$ and in fig.1b E(t) as
a function of the reduced parameter $(\tau-t)/\tau$ (notice that
because of the constancy of F the only control parameter is time).
In agreement with experiments\cite{energy,noiprl,noi}
 we find that $\epsilon$  is   power law
distributed  and that the cumulated energy $E$ has a power law
dependence on $(\tau-t)/\tau$ for $t\rightarrow \tau$. More
details on this statistical features of the model for different
driving forces will be given in a longer report. Here we want to
focus on  failure time $\tau$ of the network in the case of a
creep test (constant F).

 We first keep $KT_d=0$ and and we study
the evolution towards failure for various $KT$ and $F$.
 The results
are summarized in fig.2.
 When   $KT\ne 0$, we observe
(see fig. 2a) that failure time $\tau $ as a function of $\frac
1{F^2}$ follows an exponential law  for any fixed value of $KT_d$:
\begin{equation}
\tau \sim \exp \left[ \left( \frac{F_0}F\right) ^2\right]
\end{equation}
where $F_0$ is a fitting parameter.
 Further,looking  at fig. 2b we notice that  the failure
time $\tau $ depends on thermal noise $KT$ as follows:
\begin{equation}
\tau \sim \exp \left( \frac A{KT}\right)
\end{equation}
where A is a fitting parameter.

 We notice that these results are
similar to the prediction of Pomeau's theory, that is eq.1 with
$d=2$. Because of this analogy with eq.\ref{const}, we are now
interested in studying
 the dependence of failure time $\tau $ on disorder noise (i.e. on $KT_d$):
to this aim, we have done simulations keeping the  thermal noise
variance $KT$ fixed to constant (not zero) values. Looking at
fig.3, we observe the following facts:

\begin{itemize}
\item[B1]  failure time $\tau $ decreases following a power law as
the disorder noise variance  $KT_d$ increases, that is the more
the media is heterogeneous, the smaller is failure time (see fig.
3a).

\item[B2]  as  $KT_d$ increases, the {\it absolute} difference
between failure times $\tau $ for different values of thermal
noise variance $KT$ decreases, that is failure time $\tau $
becomes less sensitive to the effective value of thermal noise, as
shown in fig.3b.

\end{itemize}

Thus one may conclude that  disorder noise  amplifies the effect
of thermal noise and reduces the dependence of $\tau $ on the
temperature; these  results allow us to  explain the recent
experimental observations. Let us recall that experiments on
microcrystals \cite{pauchard}, gels \cite {bonn} and macroscopic
composite materials \cite {noi} reveal that dependence on $P$ of
the sample life time is very well fitted by eq.(\ref{const}).
However calculations \cite{noi} show that thermal fluctuations are
too little to activate the nucleation of microcraks in the times
$\tau $ measured in the experiments. It had been measured that the
temperature needed to have the measured
life-times $\tau $ should be of the order of several thousands of Kelvin ($%
3000K$ for wood \cite{noi} ). Experiments
 show that the life time $\tau $ of very
heterogeneous materials \cite{noi}, is (in the limit of
experimental errors) independent of  $T$ while the life time $\tau
$ of quite homogeneous materials as microcristals \cite{pauchard}
strongly depends on $T$, as predicted by eq. (\ref{const}). To
explain these results has been supposed that disorder amplifies
the thermal noise so that the nucleation time of defects becomes
of the order of the measured ones and that the lifetime $\tau $ of
the sample depend on the heterogeneity of the media. These
hypothesis are now well verified by the numerical results B1 and
B2 of the fiber bundle model.

As a conclusions we have shown that  adding a white noise, which
plays the role of temperature,  to the fiber bundle model we can
reproduce the recent experimental observations on the dependence
of  sample life time on the applied stress. Furthermore using such
a simple model we can prove  that the disorder of the system
amplifies the thermal noise in crack nucleation process. This
explains quite well why the experimentally measured  $T_{eff}$ is
more close to the thermodynamic temperature in microcrystals than
in heterogeneous materials. Similar conclusions about a disordered
induced high temperature in nucleation processes  have been
reached in other disordered systems such as foams \cite{sollich}.
This is quite interesting because it seems to be a quite general
properties of disordered systems where a thermal activated
processes with long range  correlation may be present.

We acknowledge useful discussion with Y. Pomeau. One of us (R.S.)
thanks "Le Laboratoire de Physique de l'E.N.S.L." for the very
kind hospitality during his visit in Lyon.

\newpage
\begin{itemize}

\item[$(\ast)$] on leave from "Facolt\'a d'Ingegneria, Univerist\'a di
Firenze, Italy ", with a SOCRATES exchange program of the European
Community
\end{itemize}

\newpage
\begin{figure}
\centerline{\epsfysize=0.9\linewidth \epsffile{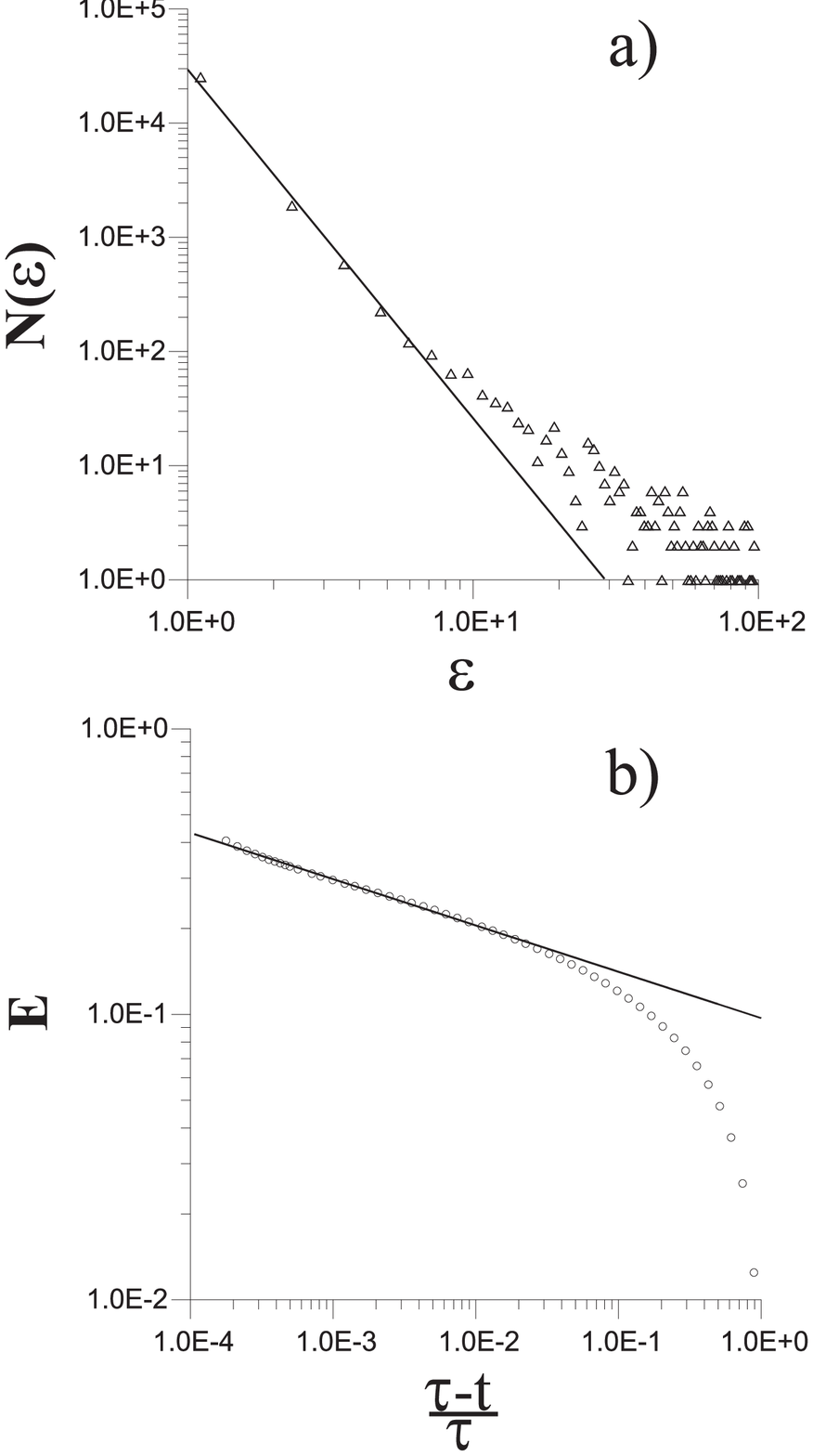}}
\caption{ (a) Distribution of energy $N(\epsilon )$ symbols  are
numerical measures, while continuous line represents a power law
fits $N(\epsilon )\sim \epsilon ^{-\beta }$ of small events with
$\epsilon<10$.  (b) $E(t)$ as a function of the critical parameter
$\frac{\tau -t}\tau $; circles are numerical measures, while the
line represent the best fit, near $t\simeq \tau $,with the
function $E(t)\sim \left( \frac{\tau -t}\tau \right) ^{-\gamma }$.
The parameters used are $KT=7 \ 10^{-3}$, $KT_d=5 \ 10^{-3}$ and
$F=540$. }
\end{figure}

\begin{figure}
\centerline{\epsfysize=0.9\linewidth \epsffile{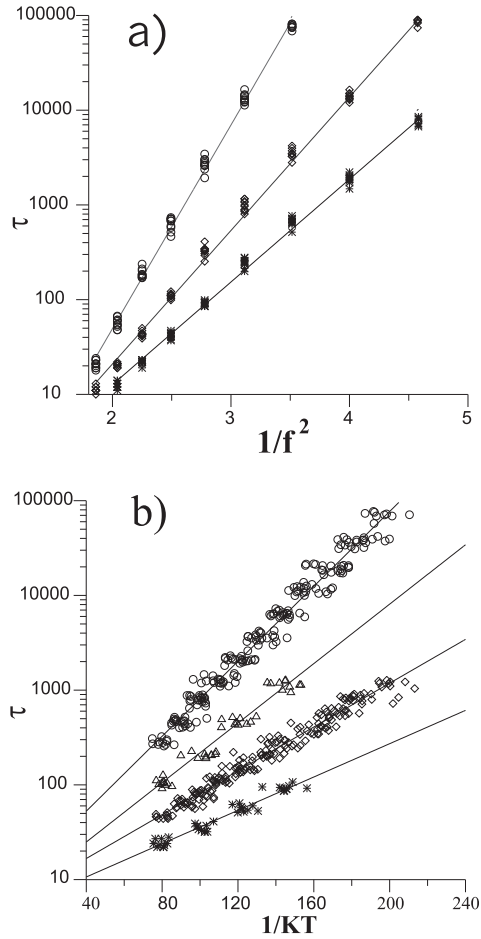}}
\caption{
 Failure time $\tau $ of an homogeneous net ($KT_d=0$) in a
creep test.
 (a) $\tau $ as a function of  the normalized force
$f=F/N$ for several values of thermal noise variance $KT$
[$(\circ) \ KT=9 \ 10^{-3}, (\lozenge)  KT=12\ 10^{-3}, (\ast)
KT=15\ 10^{-3}$]. (b) $\tau $ as a function of $KT$ for several
values of f [ $(\ast) f=0.63, (\lozenge) f=0.6, (\vartriangle)
f=0.56, (\circ) f=0.53$]. Continuous lines in (a) and (b) are  the
fits with eq.7 and eq.8 respectively. }
\end{figure}

\begin{figure}
\centerline{\epsfysize=0.9\linewidth \epsffile{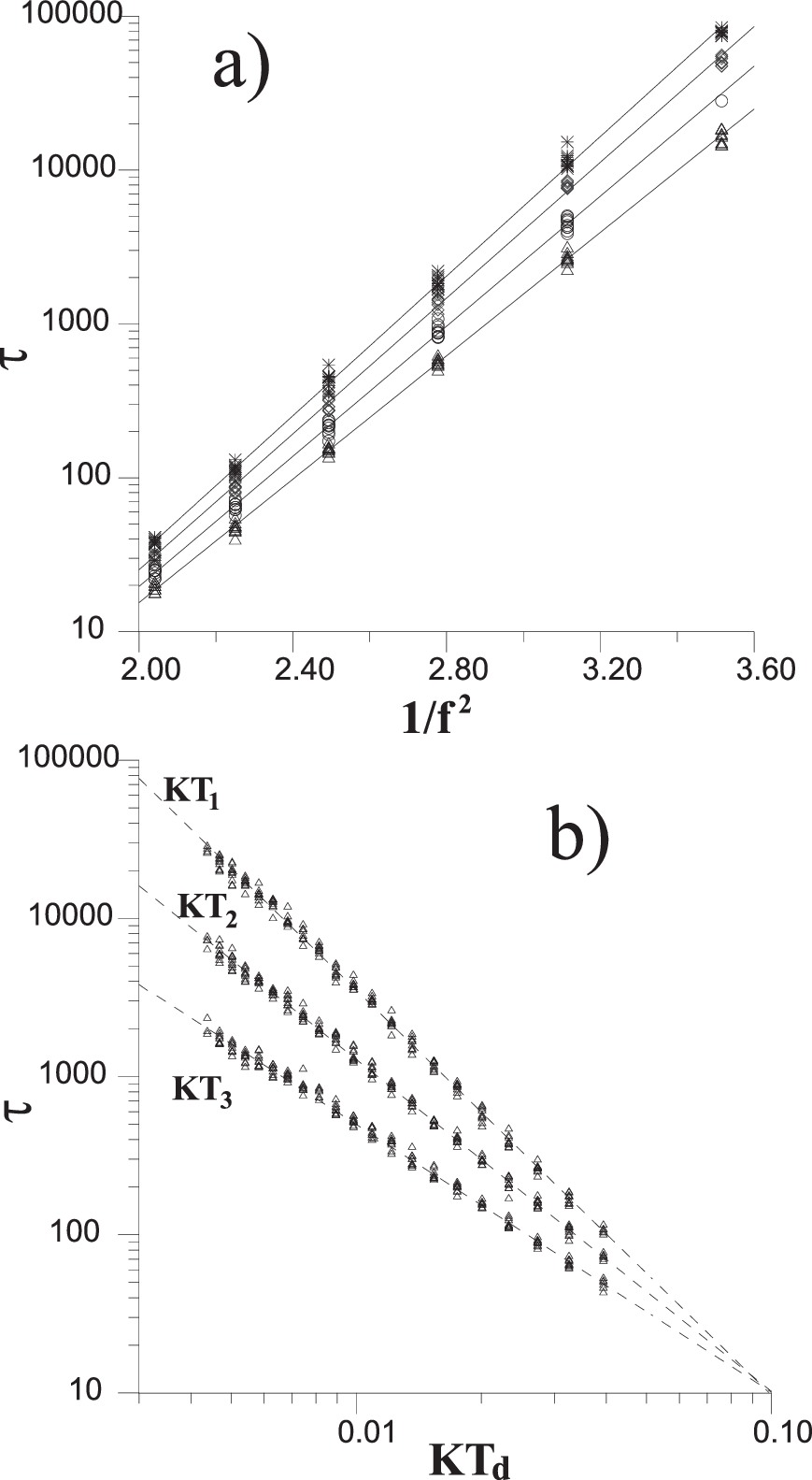}}
\caption{
  Failure time $\tau $ of an heterogeneous network ($KT_d\neq 0$)
in creep test. (a) $\tau $ is plotted as a function of normalized
force $f=F/N$, at $KT=6 \ 10^{-3}$, for different values of $KT_d$
[ $ (\vartriangle)KT_d=9.8 \ 10^{-3},  (\circ) \ KT_d=8 \ 10^{-3},
(\lozenge) KT_d=7\ 10^{-3}, (\ast) KT_d=6\ 10^{-3}$ ];
 continuous lines represent the
best fits with eq. 7.
 (b)  $\tau $ as a function of disorder noise variance
$KT_d$ at  $F=540$ for several values of  $KT$  [$ KT_1=7 \
10^{-3}, KT_2=8 \ 10^{-3}, KT_3=10^{-2}$]. Dashed lines represent
the data fits . }
\end{figure}


\begin{thebibliography}{99}
\bibitem{energy} The potential energy of a fiber which breaks is
proportional to its elastic potential energy cumulated till the
crack. If we assume that the acoustic energy is proportional to
this  potential energy then $\epsilon$ may be compared to the
acoustic energy measured in the experiments.
\end{thebibliography}
\end{document}